\documentclass{article}
\usepackage{spconf,amsmath,graphicx}
\usepackage{textcomp}
\usepackage{xcolor} 

\title{Speaker Identification using EEG}
\name{Gautam Krishna \qquad Co Tran  \qquad Mason Carnahan \qquad Ahmed H Tewfik}
\address{Brain Machine Interface Lab, The University
of Texas at Austin \\}
\begin{document}
%
\maketitle
\begin{abstract}
In this paper we explore speaker identification using electroencephalography (EEG) signals. The performance of speaker identification systems degrades in presence of background noise, this paper demonstrates that EEG features can be used to enhance the performance of speaker identification systems operating in presence and absence of background noise.  The paper further demonstrates that in presence of high background noise, speaker identification system using only EEG features as input demonstrates better performance than the system using only acoustic features as input.  
\end{abstract}
\begin{keywords}
Speaker Identification, EEG, Deep Learning
\end{keywords}
\section{Introduction}
\label{sec:intro}
Speaker identification is the task of determining an unknown speaker's identity using speaker's speech. The problem of speaker identification differs from the related problem of speaker verification or authentication. In speaker verification or authentication, if the speaker claims to be of a certain identity, then his or her voice is used to verify the claim. In this paper we study only speaker identification problem and not the verification problem. 

Recently researchers have started using deep learning models to implement speaker identification systems. The references \cite{dhakal2019near,zhao2015deep,chung2018voxceleb2,lukic2016speaker,ravanelli2018speaker,snyder2018x,ghahabi2014deep,richardson2015deep,lei2014novel,li2017deep}
explains some of the related works on deep learning based speaker identification systems. Even though deep learning models have helped to establish a new state-of-the art performance for speaker identification, their performance degrades in presence of background noise like in the case of automatic speech recognition (ASR) systems. Recently researchers have started exploring the use of physiological signals to improve the performance of automatic speech recognition systems in presence of background noise as demonstrated by the work explained in reference \cite{krishna2019speech} whether authors demonstrated that electroencephalography (EEG) signals can help isolated word based ASR systems to overcome performance loss in presence of background noise. EEG is a non invasive way of measuring electrical activity of human brain. The EEG sensors are placed on the scalp of subject to obtain EEG electrical signal readings. The EEG signals demonstrate excellent temporal resolution which makes them ideal signals to be used in brain computer interface (BCI) systems. Recently researchers have also demonstrated continuous speech recognition using EEG signals on limited English vocabulary where EEG signals are translated to English text as demonstrated by the work explained in references \cite{krishna20,krishna2019state}. 
Similarly in \cite{krishna2020synthesis} authors demonstrated preliminary results for synthesizing speech from EEG features. 

A recent work described in \cite{han2019robust} demonstrates that in fact EEG features can improve the performance of speaker verification systems operating in presence of background noise. However in \cite{han2019robust} authors did not study the problem of speaker identification. In this paper we investigate whether it is possible to improve the performance of speaker identification systems using EEG features. In \cite{krishna2019voice} authors demonstrated that EEG features are also helpful in enhancing the performance of voice activity detection (VAD) systems operating in presence of background noise. If we are able to perform speaker identification using EEG features it will also help people with speaking disabilities or people who are not able to speak to use identification systems. Face identification systems performance degrades significantly in presence of darkness. Thus if we are able to design speaker identification systems that are robust to high background noise, that will give them significant advantage over face identification systems. 

In this paper we demonstrate that EEG features can be used to enhance the performance of speaker identification systems operating in presence and absence of background noise.  The paper further demonstrates that in presence of high background noise, speaker identification system using only EEG features as input demonstrates better performance than the system using only acoustic features as input.

\section{Speaker Identification Model}
\label{sec:format}
Our speaker identification model takes EEG or acoustic or concatenation of EEG and acoustic features as input and predicts the identity of the speaker. The model architecture is described in Figure 1. The model consists of a single layer of temporal convolutional network (TCN) \cite{bai2018empirical} consisting of 128 filters followed by a single layer of gated recurrent unit (GRU) \cite{chung2014empirical} consisting of 128 hidden units followed by a dense or fully connected layer. The last time step output of the GRU layer is fed into the dense layer. 
The dense layer consists of linear activation function and number of hidden units same as total number of subjects. In this work we demonstrated our results for two data sets, the first data set had 4 subjects and second one had 8 subjects, thus in our case the dense layer can have 4 or 8 hidden units. Finally the dense layer output is passed to a softmax activation function which outputs the prediction probabilities. The labels were one hot vector encoded. 

The model was trained for 300 epochs when experimented with the first data set and for 500 epochs when experimented using the second data set. We used categorical cross entropy as the loss function with adam \cite{kingma2014adam} as the optimizer. The batch size was set to 100. The validation split hyper parameter was set to a value of 0.1. All the scripts were written using Keras deep learning framework.

\begin{figure}[h]
\label{fig:asrmodel}
\includegraphics[height=8.5cm, width=\linewidth,trim={0.1cm 0.1cm 0.1cm 0.1cm}]{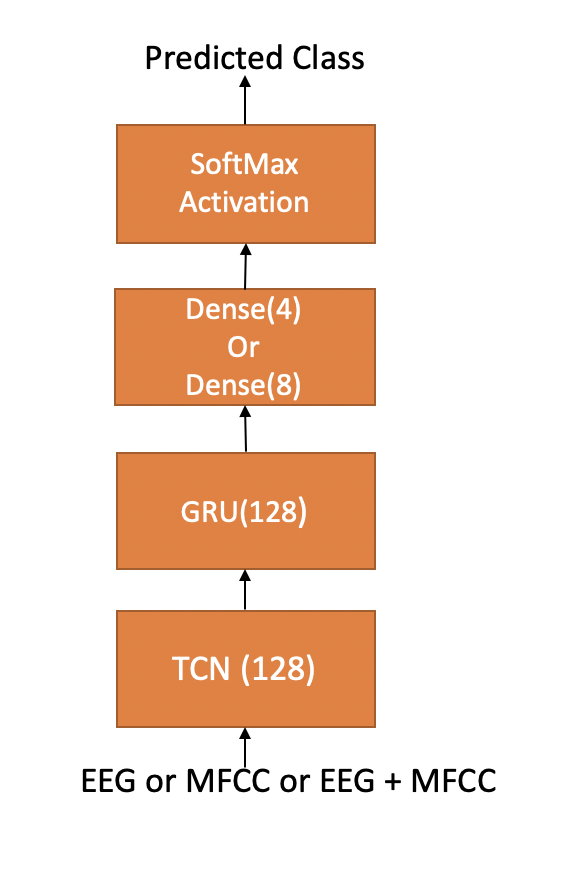}
\caption{Speaker Identification Model} 
\label{1vsall}
\end{figure}

\section{Data Sets used for performing experiments}
\label{sec:pagestyle}

We used two data sets for performing experiments. Both the data sets consists of simultaneous speech and EEG recordings. The first data set we used was the data set used by authors in \cite{krishna2020synthesis}. It consists of speech-EEG data recorded in absence of background noise for four subjects. We use only the spoken speech and EEG, ie: The EEG recorded in parallel with spoken speech, not the listening utterances or listen EEG. 
We will refer to this data set as Data set A in this work.

The second data set we used was the data set B used by authors in \cite{krishna20}. It consists of speech-EEG data recorded in presence of a background noise of 65dB for eight subjects. We will refer to this data set as Data set B in this work.

More details of the data sets, description of the EEG recording hardware used are explained in \cite{krishna20,krishna2020synthesis}. For each data set we used 80\% of the data as training set, 10\% as validation set and remaining 10\% as test set. 
A figure describing the EEG sensor locations used in the EEG cap is shown in Figure 2. 

\begin{figure}[h]
\begin{center}
\includegraphics[height=3cm,width=0.25\textwidth,trim={1cm 1cm 1cm 0.1cm},clip]{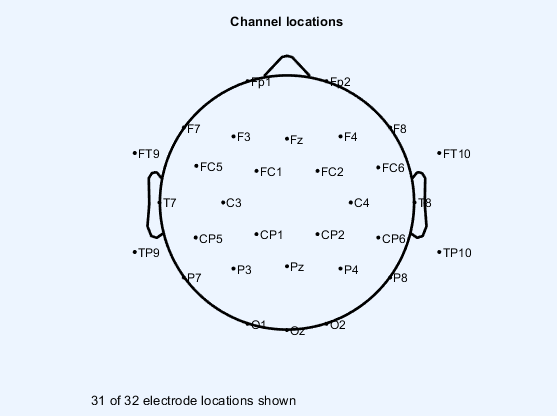}
\caption{EEG channel locations for the cap used in the experiments to collect data} 
\label{1vsall}
\end{center}
\end{figure}

\section{EEG and Speech feature extraction details}
\label{sec:typestyle}

We followed the same preprocessing methods used by authors in \cite{krishna20} to preprocess EEG and speech signal.  
The EEG signals were sampled at 1000Hz and a fourth order infinite impulse response  (IIR) band pass filter with cut off frequencies 0.1Hz and 70Hz was applied. A notch filter with cut off frequency 60 Hz was used to remove the power line noise.
The EEGlab's \cite{delorme2004eeglab} Independent component analysis (ICA) toolbox was used to remove other biological signal artifacts like electrocardiography (ECG), electromyography (EMG), electrooculography (EOG) etc from the EEG signals. 
We then extracted the five EEG features  explained by authors in \cite{krishna20}. The details of  EEG features set are covered in \cite{krishna20}. The EEG features were extracted at a sampling frequency of 100 Hz for each EEG channel. The dimension of EEG feature space was 155. 

The recorded speech signal was sampled at 16KHz frequency. We extracted mel-frequency cepstral coefficients (MFCC) as features for speech signal. We extracted MFCC features of dimension 13.
The MFCC features were also sampled at 100Hz same as the sampling frequency of EEG features. 

\section{EEG Feature Dimension Reduction Algorithm Details}
\label{sec:majhead}

We used kernel principal component analysis (KPCA) \cite{mika1999kernel} to de-noise the EEG feature space by performing dimension reduction as demonstrated by authors in \cite{krishna20}. We reduced the 155 EEG feature space to a dimension of 30 after identifying the right number of components by plotting the cumulative explained variance plot \cite{krishna2019state}.

\section{Results}
\label{sec:print}
We used classification test accuracy as the performance metric to evaluate the speaker identification model during test time. Classification test accuracy is defined as the ratio of number of correct predictions given by the model on test set to total number of predictions given by the model on test set data. Table 1 demonstrates speaker identification results obtained during test time for Data set A. As seen from Table 1 for Data set A, the highest performance was observed when the model was trained and tested using concatenation of EEG and acoustic features. 

Table 2 demonstrates speaker identification results obtained during test time for Data set B. As seen from Table 2 for Data set B, the highest performance was observed when the model was trained and tested using only EEG features but for Data set B we also observed that when the model was trained and tested using concatenation of acoustic and EEG features, it resulted in better performance than training and testing the model with only acoustic features.  

The Figure 3 shows the training and validation accuracy plot for the identification model when trained with concatenation of acoustic and EEG features for Data set A. The Figure 4 shows the training and validation accuracy plot for the identification model when trained using only EEG features for Data set B.

The overall results from Tables 1, 2 demonstrates that EEG features are helpful in enhancing the performance of speaker identification systems. We noted an interesting observation for speaker identification experiment done in presence of background noise, as seen from Table 2 the performance of the system using only EEG was better than the performance of the system using concatenation of acoustic and EEG features. One possible explanation for this observation might be that the acoustic features were extremely noisy and the model might have needed more training examples of MFCC+EEG features to achieve better generalization. Another reason might be the nature of EEG data set, in Data set A the subjects speak out loud the utterances that they listened to and their EEG signals were recorded whereas in Data set B the subjects read out loud English sentences shown to them on computer screen, in both cases the EEG signals corresponding to speech production might have different properties, this needs further understanding which will be considered for our future work. 
However we observed that in presence of background noise the speaker identification system trained with EEG features demonstrated better performance than the system trained with acoustic features. In the both the experiments ( from Tables 1, 2) \textbf{MFCC+EEG} always demonstrated better performance than only \textbf{MFCC} for speaker identification during test time.

\begin{table}[!ht]
\centering
\begin{tabular}{|l|l|l|}
\hline
\textbf{\begin{tabular}[c]{@{}l@{}}MFCC\\ (\% Test\\ Accuracy)\end{tabular}} & \textbf{\begin{tabular}[c]{@{}l@{}}EEG\\ (\% Test Accuracy)\end{tabular}} & \textbf{\begin{tabular}[c]{@{}l@{}}MFCC +EEG\\ (\% Test Accuracy)\end{tabular}} \\ \hline
45.56                                                                        & 43.33                                                                     & \multicolumn{1}{c|}{\textbf{56.11}}                                             \\ \hline
\end{tabular}
\caption{Test time results for speaker identification for \textbf{ Data set A (Absence of background noise)}}
\end{table}

\begin{table}[!ht]
\centering
\begin{tabular}{|l|l|l|}
\hline
\textbf{\begin{tabular}[c]{@{}l@{}}MFCC\\ (\% Test\\ Accuracy)\end{tabular}} & \textbf{\begin{tabular}[c]{@{}l@{}}EEG\\ (\% Test Accuracy)\end{tabular}} & \textbf{\begin{tabular}[c]{@{}l@{}}MFCC +EEG\\ (\% Test Accuracy)\end{tabular}} \\ \hline
25.69                                                                        & \textbf{59.72}                                                            & \multicolumn{1}{c|}{26.39}                                                      \\ \hline
\end{tabular}
\caption{Test time results for speaker identification for \textbf{ Data set B (Presence of background noise)}}
\end{table}

\begin{figure}[h]
\label{fig:asrmodel}
\includegraphics[height=8.5cm, width=\linewidth,trim={0.1cm 0.1cm 0.1cm 0.1cm}]{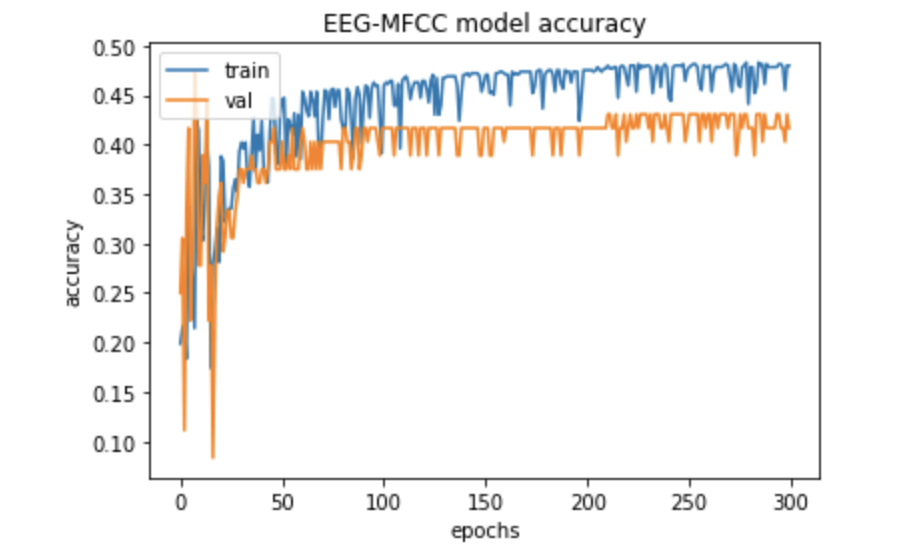}
\caption{Training and Validation accuracy for model trained with concatenation of EEG and acoustic features for Data set A \textbf{(absence of background noise)}} 
\label{1vsall}
\end{figure}

\begin{figure}[h]
\label{fig:asrmodel}
\includegraphics[height=8.5cm, width=\linewidth,trim={0.1cm 0.1cm 0.1cm 0.1cm}]{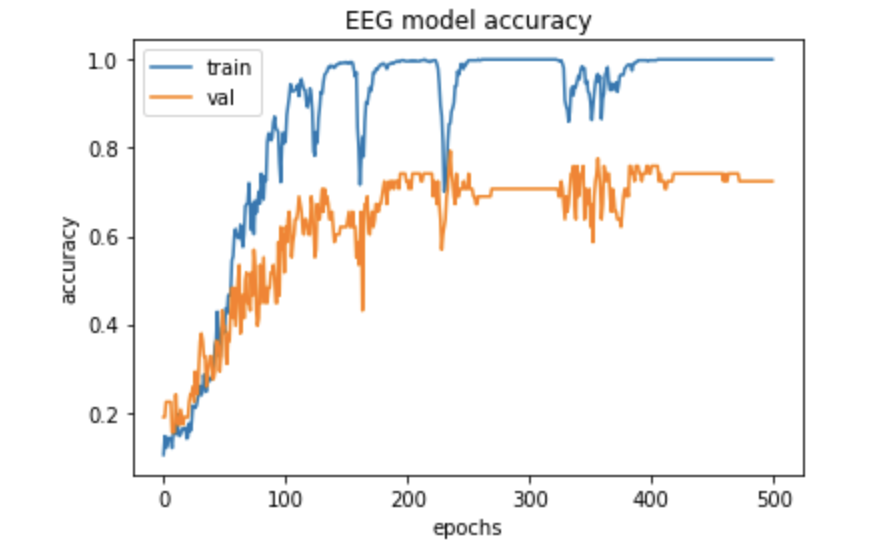}
\caption{Training and Validation accuracy for model trained with EEG features for Data set B \textbf{(presence of background noise)}} 
\label{1vsall}
\end{figure}


\section{conclusion}
\label{sec:refs}
In this paper we demonstrated that EEG features can be used to enhance the performance of speaker identification systems operating in presence and absence of background noise. We further demonstrated that in presence of high background noise, speaker identification system using only EEG features as input demonstrates better performance than the system using only acoustic features as input.

Future work will focus on improving current results by training the model with more number of examples and also validating the results on data sets consisting of more number of subjects or speakers.

\section{Acknowledgement}
We would like to thank Kerry Loader and Rezwanul Kabir from Dell, Austin, TX for donating us the GPU to train the models used in this work.
\bibliographystyle{IEEEbib}
\bibliography{strings,refs}

\end{document}